\documentclass{aa}
\usepackage{graphics}

\begin{document}

\thesaurus{07(02.16.2; 03.13.2; 08.02.2; 12.07.1)}

\title{Inversion of polarimetric data from eclipsing binaries}
\author{I J Coleman\thanks{\emph{Present address:} British Antarctic
Survey, Cambridge, CB3 0ET, UK} \and  N Gray \and J F L Simmons}

\offprints{N Gray}

\date{Received 4 February 1997 / Accepted 18 February 1998}

\institute{Department of Physics and Astronomy\\
University of Glasgow\\
Glasgow G12 8QQ\\
Scotland, U.K.}





\maketitle

\begin{abstract}

We describe a method for determining the limb polarization and limb
darkening of stars in eclipsing binary systems, by inverting
photometric and polarimetric light curves.

Because of the ill-conditioning of the problem, we use the
\relax Ba\-ckus-Gil\-bert\relax {} method to control the resolution and stability of the
recovered solution, and to make quantitative estimates of the maximum
accuracy possible. Using this method we confirm that the limb
polarization can indeed be recovered, and demonstrate this with
simulated data, thus determining the level of observational accuracy
required to achieve a given accuracy of reconstruction.  This allows
us to set out an optimal observational strategy, and to critcally assess
the claimed detection of limb polarization in the Algol system.

The use of polarization in stars has been proposed as a diagnostic
tool in microlensing surveys by Simmons et al.\ (\cite{simmons95}), and
we discuss the extension of this work to the case of microlensing of
extended sources.


\end{abstract}

\section{Introduction}

Scattered light emerging from a stellar atmosphere is expected to be
partially linearly polarized. This effect should be greatest at the
limb of the star, with a pure electron scattering atmosphere giving a
limb polarization of 11.7\% (Chandrasekhar \cite{chandra46},
\cite{chandra50}).  Less idealised calculations (Collins \& Buergher
\cite{Collins+Buerger}) suggest a lower degree of limb polarization,
around 2\% for early-type stars, though this figure is rather
sensitive to the ionization state of the outermost atmospheric layers.
Clearly a spherically symmetric star will exhibit no net polarization,
but if the symmetry is broken by an eclipse the limb polarization
should in principle be detectable. Theoretical polarimetric light
curves for this situation have been calculated by Landi Degl'Innocenti
et al.\ (\cite{landi88}).

The first detection of this `Chandrasekhar effect' was in the
\object{Algol} system (Kemp et al.\ \cite{kemp83}). This data was
analysed by Wilson and Liou (\cite{wilson93}), but the complexity of
the system and the amount of modelling involved in the analysis
prevented them from making any reliable estimate of the limb
polarization.

The inversion of polarimetric light curves from eclipsing binary stars
should allow the limb polarization of the eclipsed star to be measured
(and the photometric light curve should similarly give the limb
darkening). In fact this inverse problem is highly ill-conditioned,
and relating observations to stellar atmosphere models is therefore
far from straightforward. In this paper we investigate the practical
feasibility of determining of limb polarization by this method.

This allows us to address three closely related issues:
\begin{enumerate}
\item We develop a method of obtaining the polarization at a point on the
stellar disc, and of estimating the error on this value. This is based on
the \relax Ba\-ckus-Gil\-bert\relax {} inversion technique.
\item We determine the maximium accuracy possible in determining limb
polarization, given a number of data points and a noise level.
\item Thus, we are able to put forward an observational strategy which
should allow the best measurement of limb polarization.

\end{enumerate}

In section 2 we give a brief overview of the problem. We set out the
formalism of the \relax Ba\-ckus-Gil\-bert\relax {} method in section 3, and discuss its
suitability for the problem at hand. Section 4 contains the
calculations for the specific case of eclipsing binary stars, and
section 5 presents the results of the inversion scheme when applied to
simulated data, and the conclusions that can be drawn from these.
Section 6 considers a simplified analogue of the Algol system,
comparing the theoretical polarization profile with the best
resolution current measurements can achieve.

\section{An overview of the problem}

The radiation field across the stellar disk can be described by the
stokes parameters, which give the unpolarized and polarized intensity.
These are usually denoted by $I$, $Q$, $U$, and $V$ (these parameters
are discussed more fully in Clarke \& Grainger
(\cite{Clarke+Grainger})). Here we shall take the circular
polarization $V$ to be zero. $U$ and $Q$ give the state of linear
polarization of the radiation.  $Q$ is obtained by measuring the
difference in intensity in two perpendicular directions, and $U$ by
measuring the difference when the polarimiter is rotated through
$45^\circ$.  The degree of linear polarization may simply be written
as $ {{(Q^2 +U^2)}^{1/2}}/{I}$ and the position angle as $\frac{1}{2}
\arctan {{U}/{Q}}$. 
Under rotation of axes by $\psi$, $Q$ and $U$
simply transform as 
$U_\mathrm{new} = U_\mathrm{old} \cos 2\psi - Q_\mathrm{old} \sin 2 \psi$ 
and 
$ Q_\mathrm{new} = U_\mathrm{old} \sin 2\psi +  Q_\mathrm{old} \cos 2\psi$. 
$I$ is invariant (as is $V$).

We will assume that the light emerging from the stellar disk is
partially linearly polarized perpendicular to the radial direction.
One can calculate the total flux from the eclipsed star by integrating
the intensity $I$ over the visible part of the disc. Similarly, to
obtain the total polarized flux one simply integrates the Stokes
parameter $Q$ over the same part of the disc, introducing a rotation
factor to transform the polarized intensity at each point into the
appropriate reference frame.
\relax \begin{equation}\label{e:Flux_I}
     F_\mathrm{I}(t) =  \frac{1}{R^2}
	\relax \int\!\!\int\relax _\mathrm{disc}\,I(r)\, A(r,\phi ,t )r
	\, \relax {\rm d}\relax  r  \relax {\rm d}\relax \phi
\relax \end{equation}\relax 
\relax \begin{equation}\label{e:Flux_Q}
	F_\mathrm{Q}(t) =  \frac{1}{R^2}
	\relax \int\!\!\int\relax _\mathrm{disc}\,Q(r)\,A(r,\phi ,t )\, \cos{2\phi }r 
	\,\relax {\rm d}\relax  r \relax {\rm d}\relax \phi
\relax \end{equation}\relax 
\relax \begin{equation}\label{e:Flux_U}
	F_\mathrm{U}(t) = \frac{1}{R^2}
	\relax \int\!\!\int\relax _\mathrm{disc}Q(r)\,A(r,\phi ,t )\, \sin {2\phi }r 
	\,\relax {\rm d}\relax  r \relax {\rm d}\relax \phi     
\relax \end{equation}\relax 
where $(r, \phi)$ are polar coordinates on the stellar disk, $R$ is
the distance from the observer to the star, and $A(r,\phi ,t )$ is
zero for an occulted point on the disk, 1 otherwise.

In this paper we address the problem of extracting $I(r)$ and $Q(r)$
from measurements of $F_\mathrm{I}$, $F_\mathrm{Q}$ and $F_\mathrm{U}$
during an eclipse. One might attempt to do this by modelling $I(r)$
and $Q(r)$ for the stellar atmosphere, performing the integrals above
and using the data to fit the free parameters in the stellar
atmosphere model. But the problem must be treated more carefully. It
is a general property of inverse problems, such as the present case,
that a very large range of source models are consistent with a given
data set. This is essentially due to the smoothing properties of the
kernel (here, the occultation function $A$). Consequently, a
forward-modelling approach can give apparently accurate, but actually
highly misleading, results. A thorough treatment of inverse problems
in astronomy can be found in Craig \& Brown (\cite{Craig+Brown}); here
we will simply state that meaningful results can only be obtained by
controlling the instabilities caused by discrete, noisy data. We have
used the \relax Ba\-ckus-Gil\-bert\relax {} inversion technique to achieve this control.
The principal advantage of the \relax Ba\-ckus-Gil\-bert\relax {} method from our point of
view is that it allows us a qualitative understanding of, and thus an
explicit quantitative control over, the compromise between bias and
stability in our inversion.

\section{The Backus-Gilbert method}

The \relax Ba\-ckus-Gil\-bert\relax {} method is one of a family of methods for attacking
inverse problems.  There are several general introductions to the
method (Parker (\cite{parker77}) gives an excellent review, and Loredo
\& Epstein (\cite{loredo89}) gives an example in an astrophysical
context) -- we give a minimal introduction here, partly in order to
fix our notation.  There is a more formal discussion of the method in
an appendix, and a thorough treatment can be found in Backus
\& Gilbert (\cite{b+g70}).

We wish to recover an \emph{underlying function} $u(r)$, in this case,
either an intensity, or the run of polarization with radius; the
parameter $r$ here is the projected radial distance from the centre of the
eclipsed star's disk, in units where the star's radius is $r=1$.  We
cannot measure this underlying function directly, but instead measure
an integral of it, $F(\vec s)$, which might be the observed luminosity or
polarisation of the eclipsed star.  This is related to the
underlying function through
\relax \begin{equation}\label{e:FuK}
	F({\vec s}) = \int_0^1 u(r) K(r;{\vec s}) \,\relax {\rm d}\relax  r,
\relax \end{equation}\relax 
where $\vec s$ is the vector between the centres of the star and its
occultor, and the kernel $K(r;\vec s)$ can be calculated \emph{a priori}.
The set of observations that we make, $f_i\equiv F({\vec s}_i)$,
is therefore related to this underlying function by
\relax \begin{equation}\label{e:fuK}
	f_i = \int_0^1 u(r) K_i(r)\,\relax {\rm d}\relax  r + n_i,
\relax \end{equation}\relax 
where $K_i(r)\equiv K(r;{\vec s}_i)$ and $n_i$ is a random admixture of
noise.  {From} these measurements we wish to
produce an estimator $\hat u(r)$ of the underlying function $u(r)$.

For the \relax Ba\-ckus-Gil\-bert\relax {} method, we suppose that the mean of the
estimator and the underlying function are related by an
\emph{averaging kernel} $\Delta(r,r')$, through
\relax \begin{equation}\label{e:deltahat}
	\relax E\left(\hat u(r)\right) = \int_0^1 \Delta(r,r') u(r') \,\relax {\rm d}\relax  r'\,.
\relax \end{equation}\relax 
Since we do not know the underlying function, the averaging kernel is
of no use to us directly; however we can study its properties, and use
our data~$f_i$ in such a way as to optimise those properties, and so
minimise the dependence of the estimate $\hat u(r)$ on the underlying
function and the noise. It is clear from \relax Eqn.~(\ref{e:deltahat}) that $\hat
u(r) = u(r)$ if $\Delta(r,r')$ is the Dirac delta function.
However, such a solution is highly sensitive to noise in the data and
hence very unstable. We attain stability by increasing the width of
$\Delta(r,r')$, and hence smoothing the recovered value over a
wider region of the source -- that is, we minimise the width of
$\Delta(r,r')$ subject to the conflicting demand that the
resolution of $\Delta(r,r')$ is sufficient for the problem at
hand, in a sense we shall make more precise below.

We relate our data $f_i$ to our estimator $\hat u (r)$ through a set of
\emph{response kernels} $q_i(r)$, which produce an estimate of the
underlying function through
\relax \begin{equation}\label{e:uqf}
	\hat u(r) = \sum_i q_i(r) f_i.
\relax \end{equation}\relax

If we substitute \relax Eqn.~(\ref{e:fuK}) into \relax Eqn.~(\ref{e:uqf}), and \emph{assume}
$\relax E\left(\sum_i q_i(r) n_i\right)=0$, then we can compare with
\relax Eqn.~(\ref{e:deltahat}), and obtain
\relax \begin{equation}\label{e:deltaqK}
	\Delta(r,r') = \sum_i q_i(r) K_i(r').
\relax \end{equation}\relax 
This allows us to form some measure
of the \emph{width} of $\Delta(r,r')$ such as
\relax \begin{eqnarray}\relax 
	\relax \mathcal{A}&\equiv
		&\int (r-r')^2[\Delta(r,r')]^2\,\relax {\rm d}\relax  r',\label{e:Awidth}\\
		&=& \sum_{ij} q_i(r) W_{ij}(r) q_j(r)    \nonumber\\
		&=& {\vec q}(r)^{T} {\vec W}(r) {\vec q}(r), \nonumber
\relax \end{eqnarray}\relax 
which depends on $q_i$, and depends on $K_i$ through the definition
\relax \begin{equation}\label{e:Wdef}
	W_{ij}(r) \equiv \int_0^1 (r'-r)^2 K_i(r') K_j(r')\,\relax {\rm d}\relax  r'.
\relax \end{equation}\relax 
This is the standard definition of the width;
others are reasonable, and may be preferable in different
circumstances.

Again assuming that $\relax E\left(\vec q\cdot\vec n\right)=0$, we can also form a
measure of the stability of \relax Eqn.~(\ref{e:uqf})
\relax \begin{equation}\label{e:Bvar}
	\relax \mathcal{B}\equiv\relax \mathop{\rm{Var}}\relax \hat u(r) = {\vec q}(r)^{T} {\vec S} {\vec q}(r),
\relax \end{equation}\relax 
which depends on $q_i$ and the noise covariance matrix
$S_{ij}\equiv\relax E\left(n_i n_j\right)$.  In our simulations below, we 
take the $n_i$ to be independent and Gaussian with standard
deviation~$\sigma$; in this case, $S_{ij}=\delta_{ij}\sigma^2$, and
our assumption $\relax E\left(\vec q\cdot\vec n\right)=0$ is true.

Finally, the demand that $\Delta(r,r')$ in \relax Eqn.~(\ref{e:deltahat}) have
unit area, leads to the constraint
\relax \begin{equation}\label{e:qR}
	{\vec q}(r) \cdot{\vec R} = 1,
\relax \end{equation}\relax 
where $R_i\equiv\int K_i(r)\,\relax {\rm d}\relax  r$.

The \relax Ba\-ckus-Gil\-bert\relax {} method consists of finding those $q_i(r)$ which
minimise
\relax \begin{eqnarray}\relax 
	\relax \mathcal{A}+\lambda\relax \mathcal{B }&=&
	    {\vec q}(r) \cdot[{\vec W}(r) + \lambda {\vec S}]\cdot {\vec q}(r)
			\nonumber \\
	&=&
	\int(r-r')^2[\Delta(r,r')]^2\,\relax {\rm d}\relax  r' + \lambda\relax \mathop{\rm{Var}}\relax \hat u(r),
		\nonumber
\relax \end{eqnarray}\relax 
for some selected parameter~$\lambda$, subject to the constraint
${\vec q}\cdot{\vec R}=1$. The minimisation problem has
explicit analytic solutions
\relax \begin{equation}\label{e:qlambda}
	{\vec q}_\lambda(r) = 
	\frac{[{\vec W}(r) + \lambda {\vec S}]^{-1}\cdot{\vec R}}%
	{{\vec R}\cdot[{\vec W}(r) + \lambda {\vec S}]^{-1}\cdot{\vec R}}
\relax \end{equation}\relax 
in terms of the
parameter~$\lambda$, and these different solutions, when combined with the
data~$f_i$ using \relax Eqn.~(\ref{e:uqf}), give different estimators $\hat
u_\lambda(r)$. The nature of the trade-off in the minimisation is
clear: in order to improve the stability of the recovery, we choose a
$\lambda$ which makes $\Delta(r,r')$ broader, and so generate
response kernels $q_i$ which extend the weighted average over a greater
number of the data points~$f_i$.  The cost of this is that the
estimate of the recovered point will be biased by the inclusion of the
extra data, and this will be more marked when the underlying function
is rapidly varying.

The first important point about the \relax Ba\-ckus-Gil\-bert\relax {} method is that the
parameter~$\lambda$ allows us to adjudicate between the conflicting
demands of minimising the width of the kernel $\Delta(r,r')$ and
minimising the sensitivity of the recovered value (which is a
realisation of the statistical variable $\hat u_\lambda(r)$) to the
measurement noise, and that this adjudication can be done \emph{prior
to any data being collected}, based only on the characteristics of the
kernel $K(r;{\vec s})$ and the noise.

Secondly, we must emphasise that the ${\vec q}_\lambda(r)$ we obtain
gives us, through \relax Eqn.~(\ref{e:uqf}), a single point in the recovered
function, $\hat u_\lambda(r)$.  This means that in this simplest
version of the \relax Ba\-ckus-Gil\-bert\relax {} method we must perform the inversion for
each value of~$r$ for which we wish to find $\hat u_\lambda(r)$.
Since the calculation of the coefficients ${\vec q}_\lambda(r)$ involves
a matrix inversion, which is an $n^3$ procedure, it can be
computationally expensive, but this limitation is acceptable in our
particular case, as we only wish to recover the polarisations at the
limb, $r=1$.  This feature has the compensation that we can if
necessary select a different optimal value of~$\lambda$ for each
recovered point.

\section{Kernels for the eclipsing case}

\begin{figure}
\resizebox{\hsize}{!}{\includegraphics{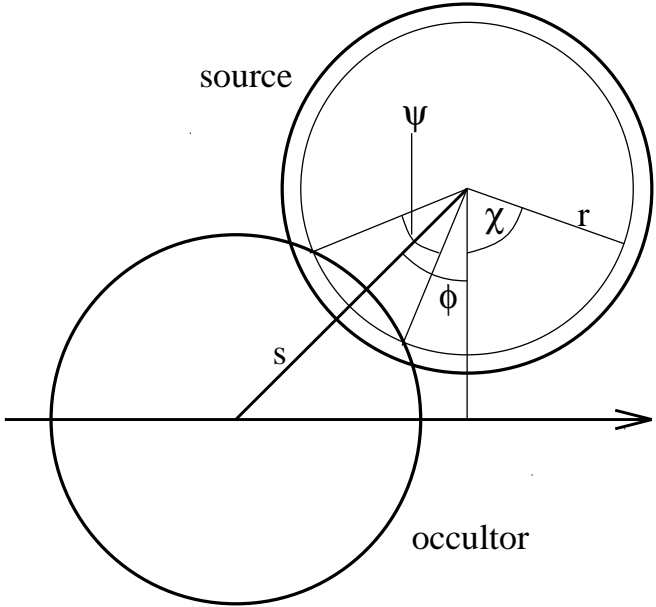}}
\caption[]{The eclipsed star has radius~1, and the
occultor radius~$\rho$.  $(r,\chi)$ are polar coordinates in the
projected plane of the source, and $(s(t),\phi(t))$ are the
coordinates of the centre of the occultor; both $\chi$ and $\phi$ are
taken from the radius which touches the point of closest approach.
When the centres of the two stars are a
distance~$s$ apart, the occultor cuts off an angle~$\psi(r;s,\rho)$ of an
annulus of radius~$r$, centred on the eclipsed star.}
\label{f:geom}
\end{figure}

In this section we derive the kernels for the case when one star is
partially eclipsed by another (a future paper (Coleman et al.\
\cite{CGS2}) will discuss the case of a gravitational lens).  The
geometry of this situation is as shown in \relax Fig.~\ref{f:geom}.
We can take the occultor to be opaque, so
that its `transfer function' $A_0(r_0)$, where~$r_0$ is the
(projected) radius from the centre of the \emph{occultor}, is
\relax \begin{equation}\label{e:Arstep}
	A_0(r_0) = \cases{0&for $r_0<\rho$\cr 1&for $r_0>\rho$}.
\relax \end{equation}\relax 
This makes $A(r,\chi;s)$, the kernel in terms of~$r$, a function of
$\psi(r;s)$.  In terms of $i(r)$, the intensity of the star as a
function of (projected) radius, the total flux from the eclipsed
star is
\relax \begin{equation}\label{e:F_I}
	F_\mathrm{I}(s) = \int_{\mbox{\scriptsize area}} i(r) A(r,\chi;s)
		\,(\relax {\rm d}\relax ^2{\vec r} = r\,\relax {\rm d}\relax  r\,\relax {\rm d}\relax \chi).
\relax \end{equation}\relax 
Thus define $\tilde A(r;s)=r\int_0^{2\pi} A(r,\chi;s)\,\relax {\rm d}\relax \chi$, so
that
\relax \begin{equation}\label{e:Idef}
	F_\mathrm{I}(s) = \int_0^1 i(r) \tilde A(r;s)\,\relax {\rm d}\relax  r,
\relax \end{equation}\relax 
putting \relax Eqn.~(\ref{e:F_I}) into the form of \relax Eqn.~(\ref{e:FuK}).

It is easy to see that $\tilde A(r;s)=r\int_0^{2\pi-\psi(s)}
1\,\relax {\rm d}\relax \chi$.  Writing $\gamma\equiv\cos[\psi(r;s)/2]$ we find that
\relax \begin{equation}\label{e:cospsi}
	\gamma = \cases{
		1&	, $\phantom{-}1<\bar\gamma$\cr
		\bar\gamma\equiv(r^2+s^2-\rho^2)/{2rs} &
			, $-1\le\bar\gamma\le1$\cr
		-1&	, $\phantom{-1\le}\bar\gamma<-1$
		}
\relax \end{equation}\relax 
We can now write
\relax \begin{equation}\label{e:tildeAdef}
	\tilde A(r;s)  = \cases{2r(\pi-\psi/2) & {} \cr
	= 2r\arccos(-\gamma)
		& , $|s-\rho|\le r \le s+\rho$\cr
	0	& , $r<-(s-\rho)$\cr
	2\pi r	& , otherwise
	}
\relax \end{equation}\relax 
defined for $r\ge 0$.

The calculation is a little more intricate for the Stokes parameters.
The light from each point on the star's disk must be linearly
polarized in the tangential direction.  Using the angle~$\chi$ defined
in \relax Fig.~\ref{f:geom}, the Stokes
parameters must therefore be
$F_\mathrm{U}(r,\chi)=-P(r)\sin 2\chi$ and
$F_\mathrm{Q}(r,\chi)=-P(r)\cos 2\chi$, for some function
$P(r)$ which we wish to recover (note that we use the unnormalised
Stokes parameters, since the normalised ones have contributions to the
noise from the intensity as well as the polarization measurements).  Defining
\relax \begin{equation}\label{e:AQdef}
	\tilde A_\mathrm{Q}(r;s,\phi)=
	-r\int_0^{2\pi}\cos2\chi A(r,\chi;s,\phi)\,\relax {\rm d}\relax \chi,
\relax \end{equation}\relax 
we therefore find that the total polarized flux in the Q direction,
measured when the centres are a distance~$s$ apart, is
\relax \begin{equation}\label{e:QPA}
	F_\mathrm{Q}(s(t),\phi(t)) = 
	\int_0^1 P(r) \tilde A_\mathrm{Q}(r;s,\phi)\,\relax {\rm d}\relax  r,
\relax \end{equation}\relax 
and similarly for $A_\mathrm{U}(r,\chi;s,\phi)$ and
$F_\mathrm{U}(s,\phi)$.  This is now in the form of \relax Eqn.~(\ref{e:FuK}).
Setting $l=s\cos\phi$, 
we can thus see that
\relax \begin{eqnarray}\relax 
	\tilde A_\mathrm{Q}(r;s,\phi)
		&=& r\cos2\phi(t)\sin\psi(r;s(t)) \nonumber\\
		&=& 2r \left(2\frac{l^2}{s^2} - 1\right)
			\gamma\sqrt{1-\gamma^2}
		\label{e:A_Q}\\
	\tilde A_\mathrm{U}(r;s,\phi)
		&=& r\sin2\phi\sin\psi \nonumber\\
		&=& 4r \frac{l}{s}\left({1-\frac{l^2}{s^2}}\right)^{1\over2}
			\gamma\sqrt{1-\gamma^2}
		\label{e:A_U}
\relax \end{eqnarray}\relax

These kernels are broad and smooth, hence the ill-conditioning of the
inverse problem.

\section{Polarization data inversion}

One strength of the \relax Ba\-ckus-Gil\-bert\relax {} method is that we can do a lot of the analysis
before we have any data.  This gives us an understanding of the
limitations of our analysis, and allows us to pick an optimal value
for the smoothing parameter~$\lambda$.

\subsection{Theoretical limits on polarization inversions}

The \relax Ba\-ckus-Gil\-bert\relax {} method is one of many inverse
problem methods in which one functional of the recovered solution,
$\relax \mathcal{A}$, is minimised subject to \emph{regulation} by another
functional $\relax \mathcal{B}$.  In this case, functional $\relax \mathcal{A}$ is the
width of the averaging kernel, and this is regulated by the
variance of the estimate.  In other inverse problem methods,
the functional $\relax \mathcal{A}$ is some measure of the goodness of fit
between the data and the forward problem, such as a $\chi^2$,
regulated by the demand that the solution be smooth, or that some
non-linear functional of the solution, such as its negentropy, be
minimised.

The \relax Ba\-ckus-Gil\-bert\relax {} scheme produces a one-dimensional family of solutions,
parametrised by the smoothing parameter $\lambda$. We can represent this as
a solution curve on a graph of standard deviation against of the recovered
$\hat u$ against the chosen measure of the width of the resolution
function: the resolution is inversely related to the width, so such a
curve illustrates the trade-off between accuracy and bias in the recovered
solution.
\begin{figure*}
\resizebox{\hsize}{!}{\includegraphics{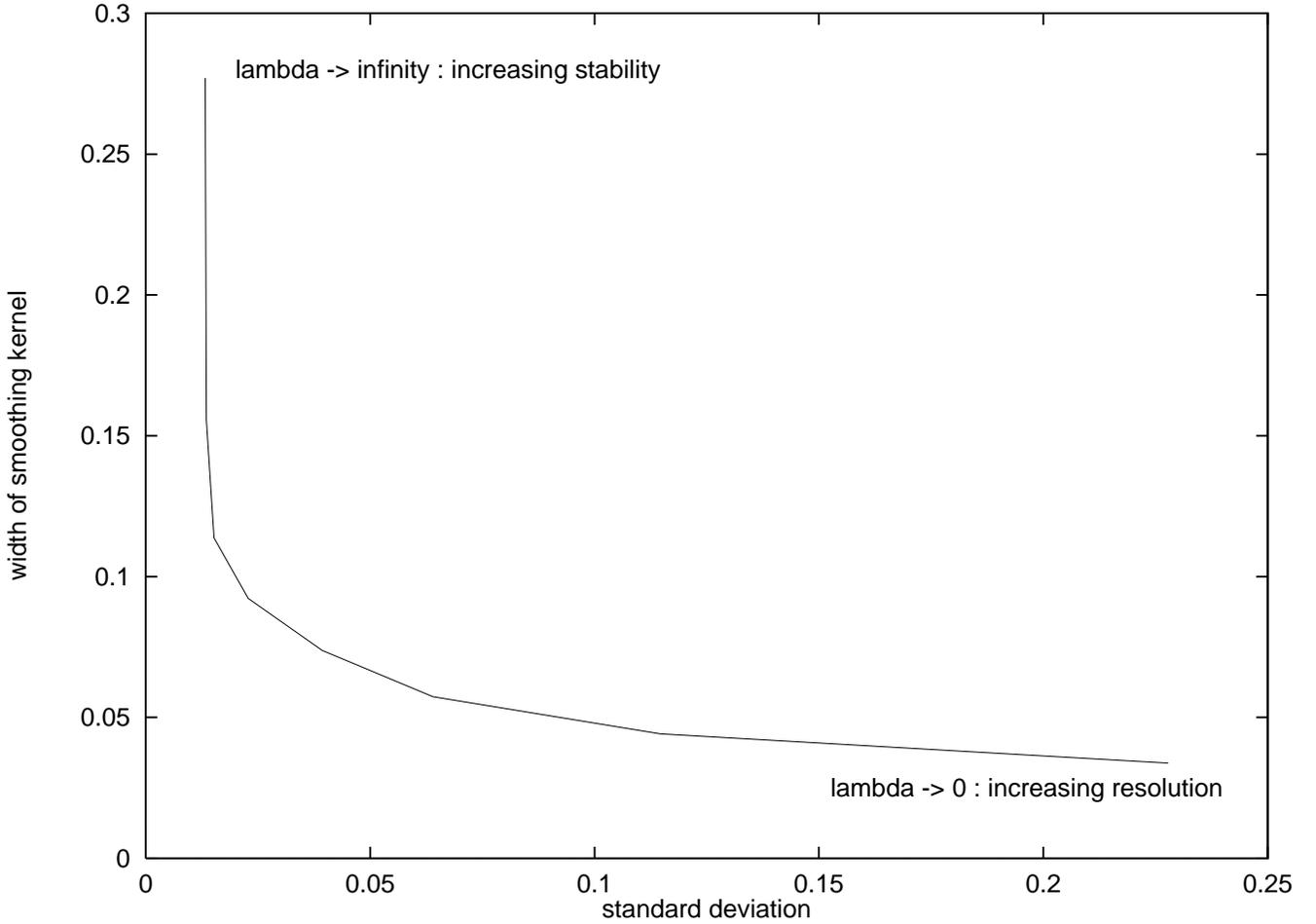}}
\caption[]{An example of a trade-off curve for the
\relax Ba\-ckus-Gil\-bert\relax {} inversion scheme.}
\label{f:tradeoff}
\end{figure*}

In general, one usually considers intermediate values of $\lambda$, around
the turning point between the two extremes. Precisely which $\lambda$ one
chooses depends on the relative importance of stability and resolution in
the particular problem.

In the present case, we are interested in measuring the polarization at the
limb, and it is clear from Fig. 2 that the resolution we need can only be
obtained at the cost of a high standard deviation in the recovered value.
If we try to increase the accuracy, we end up no longer measuring the limb
polarization proper, but rather a ``blurred'' value of the polarisation,
smoothed by convolution with the averaging kernel. If the polarization is a
maximum at the limb, the bias introduced by this smoothing will reduce the
estimated polarisation: the more sharply the polarization falls away from
the limb, the greater this biasing will be.

Looking at the low-$\lambda$ (ie, no-noise)
limit, we find that the $\Delta(r,r')$ function is sharply peaked
at $r=1$, so that in principle we can extract a well-resolved
limb-polarization.  In fact, the quality of this peak degrades
substantially for $r<1$: the method as presented \emph{cannot}
reasonably resolve polarizations on the disk.  If we wished to recover
these, we might use our knowledge of the kernels and include in the
sum in \relax Eqn.~(\ref{e:uqf}) only those kernels which are non-zero in $[0,r]$.
Such a procedure would give no improvement for $r=1$.

\begin{figure*}
\resizebox{\hsize}{!}{\includegraphics{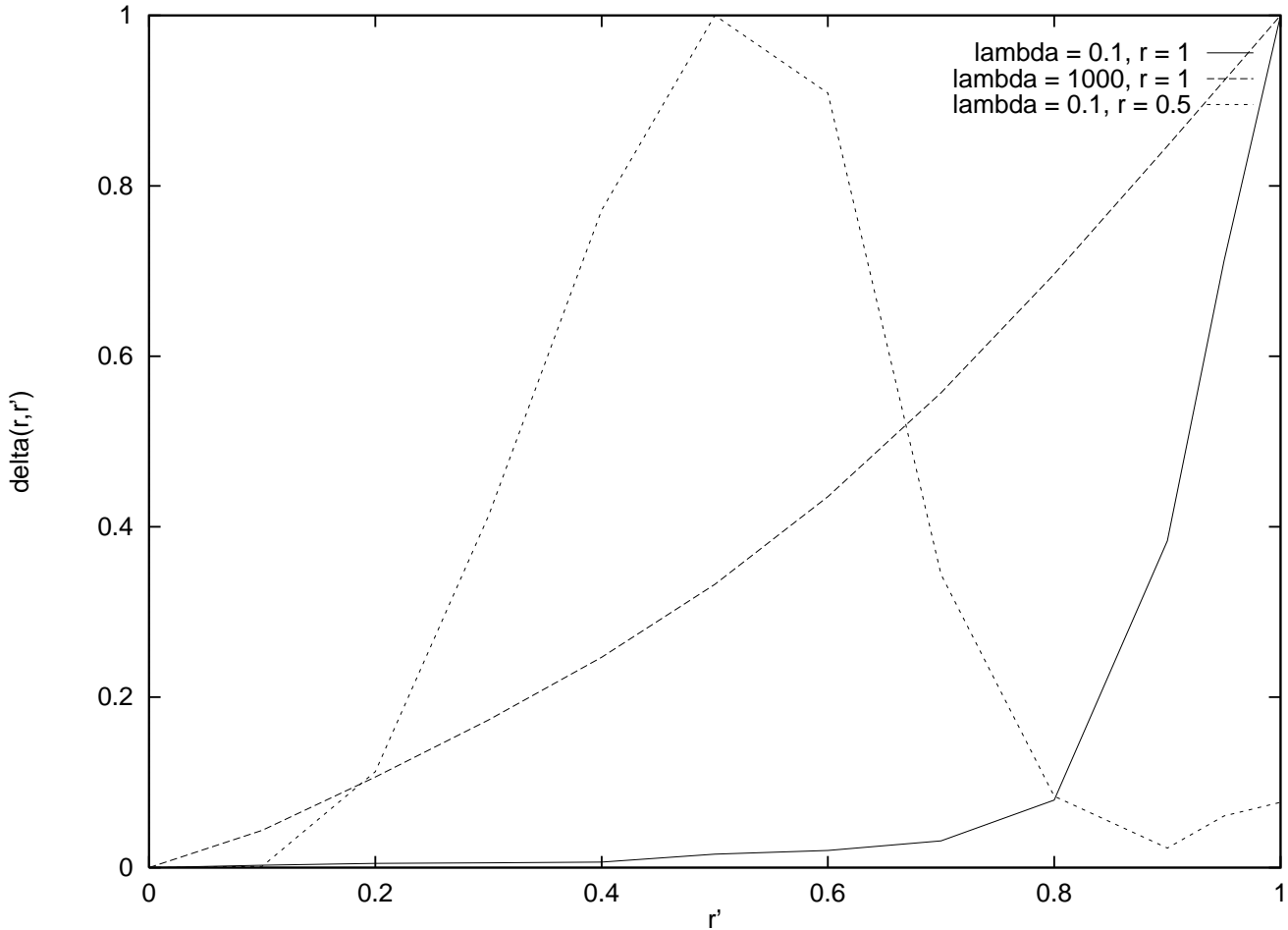}}
\caption[]{The shape of the averaging kernel
$\Delta(r,r')$ as a function of $r'$ for three different cases. The
increase in the width of $\Delta(r,r')$ as $\lambda$ increases is
clear. Note the poor resolution at $r = 0.5$, even when $\lambda$ is
small.}
\label{f:deltas}
\end{figure*}

As we increase~$\lambda$, the peak broadens, but the variance of the
recovered $\hat u_\lambda(r)$ decreases.
\begin{figure*}
\resizebox{\hsize}{!}{
\includegraphics{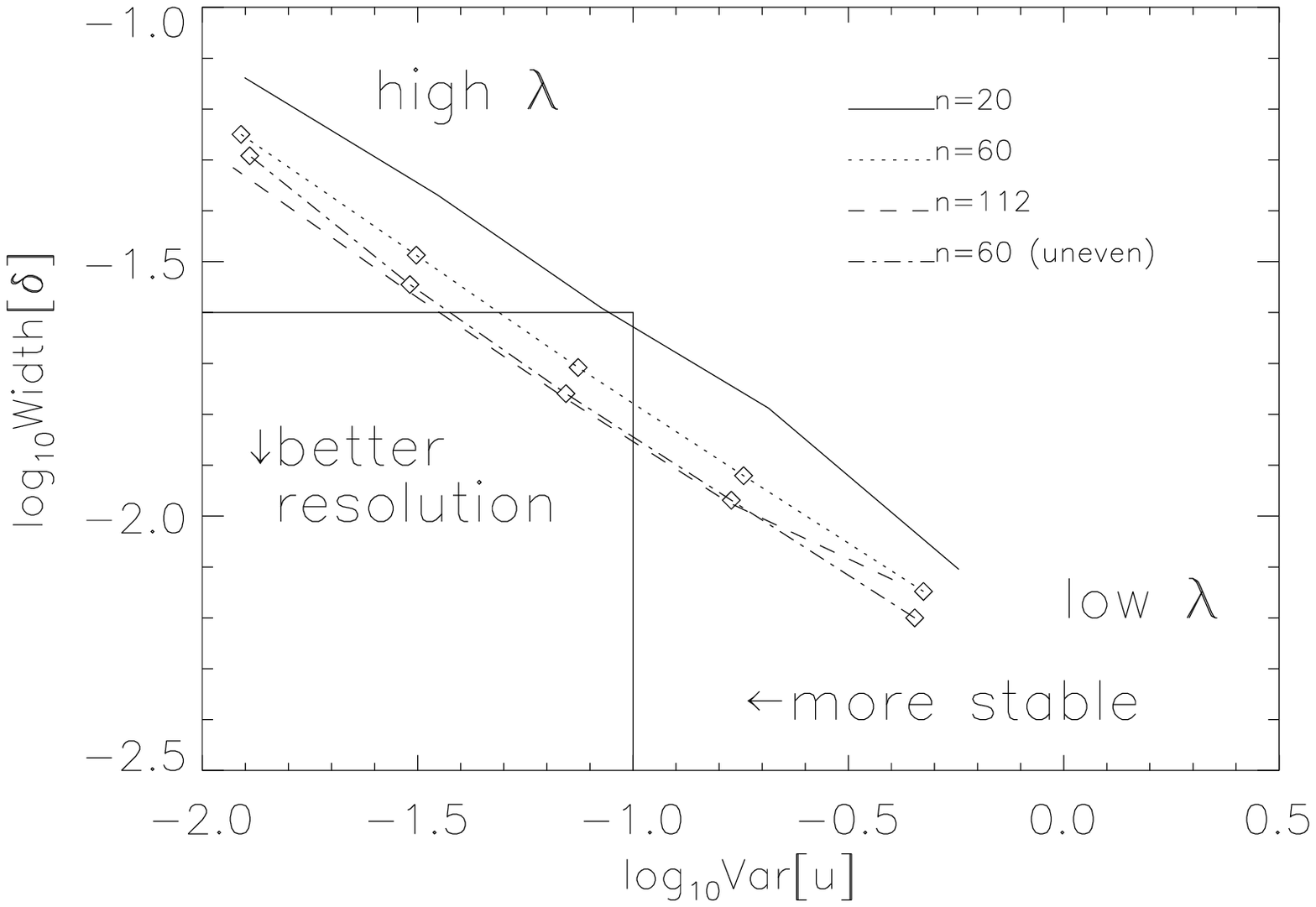}}
	\caption[]{The width of $\Delta(r,r')$,
	$\relax \log_{10}\relax \relax \mathcal{A}$, versus $\relax \log_{10}\relax \relax \mathcal{B}=\relax \log_{10}\relax \relax \mathop{\rm{Var}}\relax [\hat u]$,
	for a variety of values of~$\lambda$, and a variety of sample
	sizes~$n$.  Along each line,~$\relax \log_{10}\relax \lambda$ rises from -2 in
	the bottom right corner, to +2 in the top left, with integer
	values of~$\relax \log_{10}\relax \lambda$ plotted with a diamond on the two $n=60$
	lines. The resolution improves as we move down the plot, and
	becomes more stable as we move to the left; both the stability
	and the resolution improve as we use more data points.  These
	curves are for $n=20$, 60, 112, as indicated; for the
	discussion of the line marked `$n=60$ (uneven)', see
	Sect.~\ref{s:obs}.  The box in the bottom left encloses those
	combinations of~$\lambda$ and~$n$ which will produce an
	estimator~$\hat u$ which is adequately stable and well-resolved.}
\label{f:lambdas}
\end{figure*}
In \relax Fig.~\ref{f:lambdas} we display the width~$\relax \mathcal{A}$ and
variance~$\relax \mathcal{B}$ we obtain when we recover~$P(1)$ using the
kernel~$A_\mathrm{Q}$, for various values of $\lambda$ from $10^{-2}$ to
$10^{2}$.  Here we can clearly see the trade-off between the
well-resolved but unstable recovery in the bottom right corner, and
the stable but badly-resolved recovery in the top left.

This diagram in a sense completes the \relax Ba\-ckus-Gil\-bert\relax {} analysis of the
inverse problem, as represented by the kernel in \relax Eqn.~(\ref{e:A_Q}), and we are
now in a position to move on to invert real or simulated data. Before we
can do that, however, we must decide what value of $\lambda$ to use. To
make that decision, we must consider the level and approximate functional
form of the polarization P(r), and use this to set the scale for the
resolution and standard deviation we need to achieve. In turn, this fixes
the number of data points $n$ we require in our data and the value of the
parameter $\lambda$ we must choose in our inversion. Despite the fact that
we are invoking a particular model at this point in our analysis, we
emphasise that this introduces no \emph{practical} model dependence. We are
using an approximate model purely to help us understand what counts as
`sufficiently stable' or `sufficiently well resolved', and after this
understanding is gained the numbers we recover remain model independent
measurements, as opposed to any method of parameter fitting.

	Firstly, Chandrasekhar suggests that the limb polarization is
	of the order of $P(1)=0.1$; we therefore need a variance which
	is at least as small as this, requiring $\relax \log_{10}\relax \relax \mathop{\rm{Var}}\relax [\hat
	u]<-1$. Secondly, if we are not to have an overly biased
	result, our resolution function must be narrow compared with
	the width of the underlying function~$P(r)$.  The resolution
	we need is therefore of order $\mathop{\rm{Width}}[P(r)]$,
	with the width functional used in \relax Eqn.~(\ref{e:Awidth}).  Taking
	$P(r)\sim\exp10r$ as representative, we find we need a
	resolution better than
	$\relax \log_{10}\relax \mathop{\rm{Width}}[\Delta]=-1.6$.  Comparison with
	\relax Fig.~\ref{f:lambdas} indicates that $\lambda=1$ and $n\ga60$
	should therefore give us a satisfactory recovery of $P(1)$.
%
The maximum accuracy achievable for a given resolution (or vice
versa) can be read off from the graph.

\begin{figure*}
\resizebox{\hsize}{!}{
\includegraphics{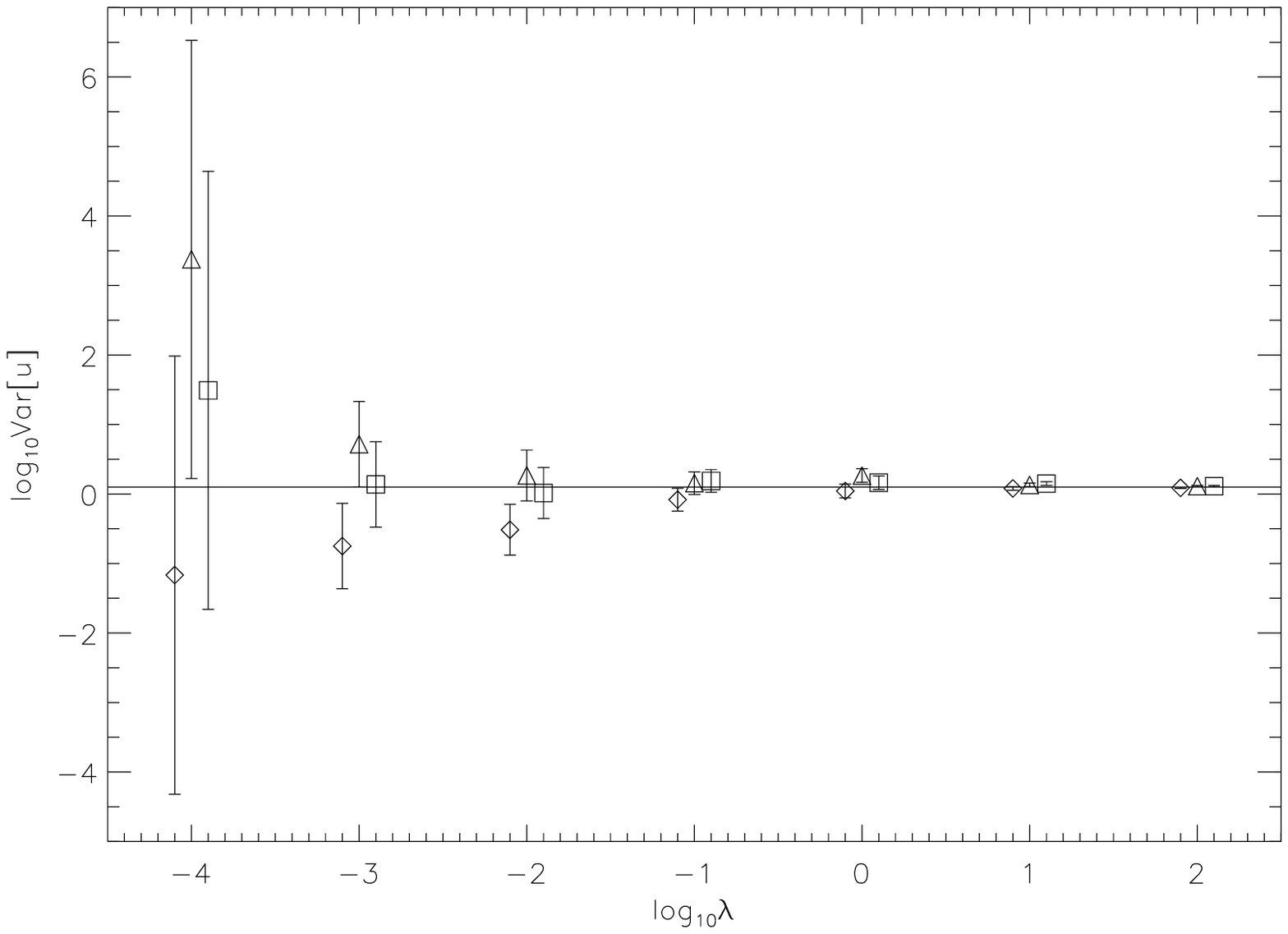}}
\caption[]{The recovered values of P(1) for three
realisations of noisy simulated data with the same parameters. The
horizontal line is the correct value of P(1).}
\label{f:recovery}
\end{figure*}

\relax Fig.~\ref{f:recovery} shows the recovery of the limb polarization from three
sets of simulated noisy data, as a function of the smoothing parameter. It
is clear that the solution becomes more stable as greater smoothing is
imposed. However, there comes a point where further smoothing does nothing
to improve the solution, but merely degrades the resolution of the
recovery.

\subsection{\label{s:alg}The Algol system: a worked example}

As a more concrete illustration, consider the \object{Algol} system.
Given the system parameters (from S\"{o}derhjelm \cite{soder}) we can
analyse the observational prospects as in the previous section. The
Algol system is more complex than the spherically symmetric cases we
consider, but our simplified analysis will provide an upper bound on
the resolution achievable with the real system.

The graphs presented by Kemp et al.\ (\cite{kemp83}) have 25 data
points in the eclipse phase. The same paper states that, in the Algol
primary star, 50\% of the polarized flux comes from an annulus on the
limb of width less than 0.005 $R_\mathrm{disc}$.

Our \relax Ba\-ckus-Gil\-bert\relax {} analysis of a spherically symmetric model system
with the Algol orbital and luminosity parameters indicates that this
limb polarization profile cannot be resolved in the Algol system. For
the eclipse coverage reported in Kemp et al.\ (\cite{kemp83}) the
minimum width $\relax \mathcal{A}_\mathrm{min}$ of the averaging kernel at the limb
is 0.0199 $R_\mathrm{disc}$, even in the zero-noise limit.

This indicates that the limb polarization cannot be truly resolved
with this data if the polarization profile is really as sharp as the
stellar models cited by Kemp et al.\ predict. Thus, while the data does
indicate the detection of limb polarization, it cannot reasonably be
used (as was attempted by Wilson and Liou (\cite{wilson93})) to make a
quantitative estimate of the polarization profile.

The precise dependence of the resolution on the number of data points
is beyond the scope of the present paper, but we note here that
increasing the number of points from 25 to as many as 1000 does not
reduce the kernel width sufficiently (from 0.019 $R_\mathrm{disc}$ to
0.0144 $R_\mathrm{disc}$). For practical purposes, then, the
inadequate resolution is intrinsic to the Algol system and will not be
alleviated by improved observations.

\subsection{\label{s:obs}Observational strategies}

Given a set of measurements~$f_i$ at positions~$s_i$, the \relax Ba\-ckus-Gil\-bert\relax {} method
can give us a well-controlled recovery of the underlying function
$u(r)$.  We can do better than this, however, as the method can
suggest how to improve our observational strategy to improve the
resolution of the recovery.  Clearly, increasing the total number of
measurements we take, or improving the noise on each measurement, will
improve the quality of our recovery.  Equally clearly, simply
binning the data (so that $\sigma\sim\sqrt n$) demonstrably
improves nothing (this can be thought of as conservation of information).

Even if we assume, however, that we cannot change the number or
quality of our data points (telescope time and equipment are limited,
after all), we might still have the freedom to adjust the
positions~$s_i$ (ie, the times) at which me make our measurements.

In \relax Eqn.~(\ref{e:uqf}), we are averaging the data points~$f_i$ with a weight
vector~$q_i$.  Where~$q_i$ is relatively large, therefore,~$f_i$ will
be smeared over several~$q_i$, or over a range of~$s_i$, smearing out
features in the kernel $K(r;s_i)$.  If we can identify prominent
features in the vector~$q_i$, and cluster our
measurements round the~$s_i$ they correspond to, we should be able to
decrease the spread of $\Delta(r,r')$ and so increase the
resolution of the recovery.

	Our simulations show that this rather informal argument is
	valid for our case.  The line in \relax Fig.~\ref{f:lambdas} captioned
	`$n=60$ (uneven)' has the same number of data points as the
	line `$n=60$', but with the values~$s_i$ chosen so that the
	`data rate' in a band $s=1.85\pm0.15$ is double that outside
	the band.  This does not significantly improve the variance of
	the result (we are not gathering any more information than
	before), but it does noticeably improve its resolution.

This improvement in our procedure corresponds to concentrating our
measurements on the point around the beginning of the eclipse.
We might have guessed that this would be a reasonable strategy to
adopt, but the \relax Ba\-ckus-Gil\-bert\relax {} method has justified our guess, and would substitute
for a lack of intuition in a more obscure situation.

\section{Conclusions}

Our studies of simulated data show the fundamental limitations on the
determination of limb polarization in eclipsing binary stars. In
particular, \relax Fig.~\ref{f:lambdas} indicates that limb polarizations of order a
few percent are only just above the threshold of detectability, even in
perfectly spherically symmetric, non-interacting binary systems. The
situation will be worse in more complex systems.

It is important to appreciate that the \relax Ba\-ckus-Gil\-bert\relax {} method does not
strictly estimate the polarization at a point on the stellar disc, but
rather the polarization convolved with the resolution function. To relate
the results of the inversion to a particular model, it is necessary to
calculate the theoretical value of this convolution, which should be
consistent with the $\lambda$ we have chosen and with all higher values of
$\lambda$ -- these represent coarser averages over the stellar disc.

The bottom line is that one must take care in drawing conclusions about
limb polarization from studies of eclipsing binaries. It is clearly not
possible to distinguish between stellar atmosphere models on this basis if
their predicted limb polarizations differ by less than the maximum accuracy
achievable. On the other hand, an appreciation of the issues raised in this
paper will allow a meaningful determination of limb polarization, with
reliable error estimates.

Formally, the eclipsing binary problem is very similar to the gravitational
microlensing problem. Indeed, part of the initial motivation for this work
sprang from studies of the microlensing of extended sources. A future paper
(Coleman et al.\ \cite{CGS2}) will apply the inverse problem approach outlined in this paper
to the use of microlensing as a probe of stellar atmospheres.

\begin{acknowledgements}

We would like to thank Dr.\ Richard K.\ Barrett for enlightening
discussions of the \relax Ba\-ckus-Gil\-bert\relax {} me\-thod. IJC was supported by a
PPARC studentship.

\end{acknowledgements}

\appendix

\section{General derivation of the Backus-Gilbert index}
\label{s:appx}

In section 3, we described the \relax Ba\-ckus-Gil\-bert\relax {} inverse in rather practical terms,
taking care to relate the description to the quantities obtained in, and
the concerns relevant to, real observations.  Such a physical
understanding of the method is essential if it is to be used
properly, but we can obtain other insights into the method by
reexamining it in a more formal way\footnote{We thank the referee, Dr
A Lannes, for suggesting this approach.}.

\relax Eqn.~(\ref{e:FuK}) above describes an operator $K: P\to D$ mapping an object from a
source space $P$ into a data space $D$.  Including noise $n\in D$, we
have
\relax \begin{equation}\label{e:fKun}
	f = Ku + n
\relax \end{equation}\relax 
for $f\in D$ and $u\in P$.  Here $P$ is a real Hilbert space,
parametrised by~$r$, with a symmetric inner product 
\[
	\relax (a|b)_{P}=\int_0^1 a(r)b(r)\,\relax {\rm d}\relax  r\quad, \forall a,b\in P,
\]
%
%
%
and $D$ is a finite-dimensional Euclidean space with
\[
	\relax (a|b)_{D }= \sum_i a_i b_i.
\]
We wish to make an estimate $\hat u_r\in \relax \bf{R}\relax $ of a single component
of the object~$u$, based on the data~$f$.  To this end, we wish to
find a $q\in D$ (depending on~$r$), such that
\relax \begin{equation}\label{e:uqfD}
	\hat u_r = \relax (q|f)_{D}.
\relax \end{equation}\relax 
We find this~$q$ as the solution of a minimisation problem.
Introducing the adjoint operator $K^*: D\to P$, and assuming 
$\relax E\left(\relax (q|n)_{D}\right)=0$, we have
\relax \begin{equation}\label{e:Eu}
	\relax E\left(\hat u_r\right) = \relax (q|Ku)_{D }= \relax (K^*q|u)_{P}.
\relax \end{equation}\relax 
Recalling that $\relax (K^*q|u)_{P}=\int_0^1(K^*q)(r')u(r')\,\relax {\rm d}\relax  r'$, we see that
$(K^*q)(r')\in P$ can be identified with the averaging kernel
$\Delta(r,r')$, and that \relax Eqn.~(\ref{e:Eu}) will be a good estimate of $u_r$
when~$q$ is transformed by~$K^*$ into the basis vector $e_r\in P$
corresponding to the component~$r$ of~$u$.  That is, \relax Eqn.~(\ref{e:Eu}) would
be exact if $K^*q=e_r$.  The object $K^*q$ will instead be a linear
combination of basis vectors `close' to~$e_r$, and we can measure its
%
%
%
%
%
`scatter' around~$e_r$ with the operator $Q:P\to P$ such that
$Qx_{r'}\equiv (r-r')^2x_{r'}, \forall x_{r'}\in P$.  Define
\relax \begin{equation}\label{e:Adef}
	\relax \mathcal{A} \equiv \relax (K^*q|QK^*q)_{P }= \relax (q|KQK^*q)_{D
		}= \sum_{i,j} q_i W_{ij} q_j,
\relax \end{equation}\relax 
defining the (self-adjoint) operator $W=KQK^*:D\to D$.  We may also
define a measure of the stability of~$\hat u_r$,
\relax \begin{equation}\label{e:Bdef}
	\relax \mathcal{B} \equiv \relax (q|Sq)_{D},
\relax \end{equation}\relax 
by analogy with \relax Eqn.~(\ref{e:Bvar}), where the operator $S\in D$ is such that
$\relax (e_i|Se_j)_{D }= S_{ij}$, where $S_{ij}$ is the positive definite
noise covariance
matrix.
The demand that $\Delta(r,r')$ have unit area translates into the
constraint $\relax (K^*q|1)_{P}=1$, where $1\in P$ is the all-1 vector in~$P$.
Writing $R\equiv K1\in D$, this is equivalent to the constraint
\relax \begin{equation}\label{e:qR1}
	\relax (q|R)_{D }= 1,
\relax \end{equation}\relax 
restricting~$q$ to a hypersurface in~$D$, with normal~$R$.

If we now introduce the functional $c: D\to\relax \bf{R}\relax $, such that
\relax \begin{equation}\label{e:cqdef}
	c(q)\equiv \frac12 [\relax \mathcal{A}(q) + \lambda\relax \mathcal{B}(q)]
		=\frac12 \relax (q|(W+\lambda S)q)_{D},
\relax \end{equation}\relax 
the minimisation problem becomes that of finding the~$q$ which
minimises~$c(q)$, subject to $\relax (q|R)_{D}=1$.
Considering  small variations~$\epsilon\in D$ in the hyperplane (that
is $\{\epsilon:\relax (\epsilon |R)_{D}=0\}$, and defining the gradient $\nabla
c(q)=(W+\lambda S)q$ ($W+\lambda S$ is self-adjoint since both~$W$
and~$S$ are), we have
\[
	c(q+\epsilon) = c(q) + \relax (\epsilon|\nabla c(q))_{D }+
		\frac12 \relax (\epsilon|(W+\lambda S)\epsilon)_{D}.
\]
This is extremised at $q_\lambda$ such that 
$\relax (\epsilon|\nabla c(q_\lambda))_{D}=0$, and is a minimum if
$\relax (\epsilon|(W+\lambda S)\epsilon)_{D }\ge 0$.
The operator~$S$ is positive-definite by definition, and the
operator~$W$ is positive-definite, since~$Q$ is.  The operator~$c(q)$
is therefore minimised when $\relax (\epsilon|(W+\lambda S)q_\lambda)_{D}=0,
\forall\epsilon$, or
\relax \begin{equation}\label{e:cqsol}
	(W+\lambda S)q_\lambda = \alpha R,
\relax \end{equation}\relax 
for any $\alpha\in\relax \bf{R}\relax $.  Imposing the constraint \relax Eqn.~(\ref{e:qR1}), we
thus find
\relax \begin{equation}\label{e:cqsol2}
	q_\lambda = \frac{(W+\lambda S)^{-1} R}%
		         {\relax (R|(W+\lambda S)^{-1)_{ }R}D},
\relax \end{equation}\relax 
from which we can obtain \relax Eqn.~(\ref{e:qlambda}), from the definition of
$\relax (\cdot|\cdot )_{D}$.

\end{document}